\documentclass[aps,prb,twocolumn,groupedaddress]{revtex4}

\begin{document}


\title{Temperature-induced barium de-trapping from a double-well potential in 
Ba$_{6}$Ge$_{25}$}


\author{M. Schmidt}
\author{P. G. Radaelli}
\author{M. J. Gutmann}
\affiliation{ISIS Facility, Rutherford Appleton Laboratory, Chilton, Didcot, Oxon OX11 0QX, United Kingdom}
\author{S.J.L. Billinge }
\affiliation{Department of Physics and Astronomy and Center for Fundamental 
Materials Research, Michigan State University, East Lansing, USA}
\author{N. Hur}
\author{S.W. Cheong}
\affiliation{Department of Physics and Astronomy, Rutgers University, Piscataway, 
New Jersey 08854, USA}


\date{\today}

\begin{abstract}
The crystal structure of barium-germanium clathrate Ba$_6$Ge$_{25}$ was studied 
using neutron powder diffraction in the temperature range 20--300K. The compound 
was found to be cubic (S.G. P4$_1$23) in the entire temperature range. However, 
the fully-ordered model of the crystal structure (no split sites) is marginal at 
room temperature, and clearly fails at low temperature. A much better 
description of the crystal structure below 250K is given in terms of two split 
Ba sites, with random occupancies, for two out of three types of cages present 
in the  Ba$_6$Ge$_{25}$ structure. The Ba atoms were found to interact strongly 
with the Ge host. The separation of the split Ba sites grows with decreasing 
temperature, with a sudden increase on cooling through the 200--250K temperature 
range, accompanied by an expansion of the entire crystal structure. 
We propose a simple model for this transition, based on temperature-induced de-
trapping of Ba from a deep double-well potential.  This transition is associated 
with sizeable anomalies in the transport and magnetic properties.  The most 
significant of these effects, that is, the drop in electrical conductivity on 
cooling, can be easily explained within our model through the enhanced 
structural disorder, which would affect the relaxation time for all portions of 
the Fermi surface.  We suggest that the other anomalies (increase in the 
absolute value of the negative Seebeck coefficient, decrease in the magnetic 
susceptibility) can be explained within the framework of the one-electron semi-
classical model, without any need to invoke exotic electron-electron interaction 
mechanisms.
\end{abstract}


\maketitle

\section{Introduction}
There is growing interest in open-structure semiconducting materials that are 
considered promising candidates for thermoelectric applications. These compounds 
are characterized by low and often glass-like thermal conductivity $\kappa$, 
high electric conductivity $\sigma$ and large Seebeck coefficient $S$.  In 
particular, a significant research effort has focused on group III and IV 
clathrates, with heavy alkali metal, alkaline earth or rare earth guests. The 
group III and IV elements form a network of cages, in which a large atom is 
hosted. Usually, the guest atom is weakly bonded to the cage \cite{blake01} and 
the sizes of cages are large enough for the atom to rattle. The semiconducting 
host can be doped to provide suitable electronic properties while the heavy 
guest atom can significantly lower the thermal conductivity of the material, due 
to resonant scattering of the heat carrying phonons 
\cite{nolas00,sales01,keppens00}, thereby increasing the thermoelectric figure 
of merit $Z=S^2\sigma/\kappa$.  These materials are a realization of the 
``phonon glass, electron crystal'' (PGEC) model proposed by Slack 
\cite{slack95}. 

The best-known materials of this class, such as (Ba, Sr, 
Eu)$_8$Ga$_{16}$Ge$_{30}$, adopt the so-called type-I clahtrate structure, which 
is also common to a variety of gas hydrates \cite{pauling52,sales01,nolas00}, 
and contains highly symmetric dodecahedral and tetrakaidecahedral cages 
\cite{chakoumakos00,sales01}.  Ba$_6$Ge$_{25}$ is also a member of the clathrate 
family, but its crystal structure and its behaviour are unusual. Ba$_6$Ge$_{25}$ 
crystallizes in the cubic P $4_1$ 3 2 space group with a lattice parameter 
$a=14.54536(7)$~\AA\ and $Z=4$.  Barium occupies three non-equivalent 
crystallographic positions in the Ba$_6$Ge$_{25}$ unit cell \cite{cabrera00,fukuoka00,kim00}.  Ba(1) (8 
equivalent sites per unit cell) is contained in distorted dodecahedral cages, 
sharing the pentagonal faces with each other and forms a spiral structure across 
the unit cell.  However, Ba(2) (4 per unit cell) and Ba(3) (12 per unit cell) 
are hosted in channel-like structures that are interconnected throughout the 
unit cell \cite{cabrera00,fukuoka00,kim00}.  Another unusual structural feature 
is the fact that 32 out of 100 Ge atoms in each unit cell are three-coordinated, 
and accommodate a lone electron \cite{cabrera00,fukuoka00,kim00}.  The 
transport and magnetic properties of Ba$_6$Ge$_{25}$ are also unusual.  Unlike 
(Ba, Sr, Eu)$_8$Ga$_{16}$Ge$_{30}$, which can be prepared in both semiconducting 
and metallic form by slightly altering the Ga/Ge ratio, Ba$_6$Ge$_{25}$ is an 
$n$-type metal, with a fairly high carrier concentration ($n\sim$1--
2$\cdot10^{22}$cm$^{-3}$), and a good room-temperature electrical conductivity 
($\sim 2000 \Omega^{-1}$cm$^{-1}$) \cite{kim00,paschen02}.  The room-temperature 
Seebeck coefficient (-20 $\mu $V/K) is also consistent with metallic rather than 
semiconducting properties.  Upon cooling below 240 K, Ba$_6$Ge$_{25}$ undergoes 
a transition, affecting the electrical conductivity (which decreases by almost a 
factor of 4), the Seebeck coefficient (which is negative and increases in 
absolute value) and the magnetic susceptibility $\chi$, (which becomes more 
diamagnetic) \cite{paschen02}.  At low temperatures, Ba$_6$Ge$_{25}$ is still a 
metal, as clearly indicated by the linear temperature dependence of the specific 
heat and Seebeck coefficient, and by the high carrier density, as determined by 
Hall effect measurements \cite{kim00,paschen02}.  Moreover, a superconducting 
transition has been discovered below 1 K \cite{grosche01}.

This intriguing behaviour is clearly very difficult to understand with the naive 
Sommerfeld approach.  In the free-electron model \cite{ashcroft}, $\sigma$ , $S$ 
and the electronic (Pauli) component of the magnetic susceptibility $\chi_e$ are 
proportional to $n$, $n\pm2/3$ and $n+1/3$, respectively.  $\sigma$ is also 
proportional to the relaxation time $\tau$, whereas the other two quantities are 
independent of $\tau$.  Qualitatively, one may conclude from the sign of the 
anomalies that a sharp reduction of the carrier concentration is occurring on 
cooling below the transition.  However, recent Hall effect and specific heat 
measurements by Paschen {\it et al.} \cite{paschen02,steglich02} indicate that 
this is not the case.  In fact, if the Hall measurements are interpreted in the 
simple single-band model, the drop in conductivity is almost entirely accounted for 
by a simultaneous drop in carrier {\it mobility} (which is proportional to 
$\tau$).  The interest in this compound has been further heightened by a report 
of a structural transition at the same temperature where the transport and 
magnetic anomalies occur.  Although the crystallographic data are still 
unpublished, they are discussed in the aforementioned work by Paschen {\em et 
al.} \cite{paschen02} and used by Zerec  {\em et al.} \cite{zerec02} to 
calculate the electronic band structure of , Ba$_6$Ge$_{25}$.  
The structural anomaly is not accompanied by a symmetry reduction. It is due to 
a strong increase of the Ba site disorder on cooling, and is consistent with the 
Ba atoms 'locking in' to two well-separated positions (split sites) at low 
temperatures.  Paschen {\it et al.} speculate that the unusual transport and 
magnetic properties of Ba$_6$Ge$_{25}$ may arise from the formation of spinless 
bipolarons, whereby two carriers with opposite spins would dynamically stabilize 
a pair of Ba atoms separated by a short nearest-neighbour distance.  Clearly, 
such a scenario may have important implications for understanding 
superconductivity in this system.

In this paper, we present detailed temperature dependent measurements of the 
Ba$_6$Ge$_{25}$ clathrate crystal structure, as determined from Rietveld 
refinements of neutron powder diffraction data in the temperature range 20--
300K.  The scattering contrast of Ge and Ba for neutrons (b$_{Ba}$=3.86 fm; 
b$_{Ge}$=8.18 fm) is reversed with respect to x-rays (Z$_{Ba}$=56; Z$_{Ge}$=32), 
providing a unique perspective not only on the large barium displacements, but 
also on the smaller displacements of the germanium framework.  We conclude that, 
at low temperatures, Ba(2) and Ba(3) are trapped in a deep double-well 
potential, and that dynamic displacement coupled with electron hopping is very 
unlikely to occur.  The trapping of barium atoms results in a dramatic 
structural rearrangement, which is bound to have a profound influence to the 
electronic states near the Fermi energy, especially those associated with narrow 
bands with predominant Ba character.  Under these circumstances, the free-
electron interpretation of the transport and magnetic properties clearly breaks 
down.  However, we find that the single-electron (band) picture in the semi-
classical approximation is most likely sufficient to describe this system, 
without any need to invoke exotic electron-electron interaction mechanisms.
\section{Experimental}
Polycrystalline Ba$_6$Ge$_{25}$ was prepared from elemental barium (99.2\% 
purity, from Alfa Aesar) and germanium (99.999\% purity, from Alfa Aesar)  mixed 
together in the molar ratio of $1.15:4$. Excess Ba was added to compensate for 
its evaporation from the reaction vessel. The mixture was placed in a closed 
graphite crucible, which was sealed in an evacuated silica ampoule. The 
reactants were slowly heated up to 1353K over a period of 10h and kept at this 
temperature for 8h. Then the sample was cooled down to room temperature over a 
period of 10h. The resulting sinter was black with brown tarnish layer on the 
surface. The brown impurity was washed away using methanol in an ultrasonic 
cleaner. The material was found to contain traces of graphite from the reaction 
crucible, as well as small traces of an unidentified phase. Neutron diffraction 
patterns were collected using the General Materials Diffractometer (GEM) at 
Rutherford Appleton Laboratory. The sample for the scattering experiment was 
enclosed in a vanadium can and attached to a closed cycle helium refrigerator. 
The measurements were carried out in the temperature range of 20--300K with 10K 
increments.  At every temperature, data was acquired for 1h at 175~$\mu$A of 
proton beam current. The Rietveld analysis of the diffraction patterns was 
carried out using the GSAS package \cite{gsas}. The crystal structure was 
visualized using ORTEP-3 program \cite{ortep}.
\section{Structure properties}
Ba$_6$Ge$_{25}$ was found to be cubic (space group P $4_1$ 3 2) in the entire 
temperature range, with no trace of additional Bragg peaks at low temperatures. 
Preliminary Rietveld refinements were carried out using a model derived from 
room temperature x-ray single crystal experiments 
\cite{fukuoka00,cabrera00,kim00}. In this paper we follow the atom labelling 
scheme of Kim {\em et al.} \cite{kim00}. 

Fig.~1 shows the $a$ lattice parameter of Ba$_6$Ge$_{25}$ as a function of 
temperature. The lattice parameter increases with increasing temperature, but 
exhibits an anomaly in the 200--240K temperature range which indicates the phase 
transition in Ba$_6$Ge$_{25}$, as reported by Paschen {\em et al.} 
\cite{paschen02}. The overall lattice expansion of Ba$_6$Ge$_{25}$ is $\Delta 
a=0.018$\AA\ ($\Delta a/a\approx 0.1$\%) in the 20--300K temperature range.  It 
should also be noted that the lattice parameter contraction associated with the 
anomaly is very small (0.004\AA). In our data the contraction happens smoothly
over a 40~K range.   This smooth cross-over is also reflected in our measurements of 
resistivity (see inset in Fig. 1) and contrasts with the sharp cross-over reported by Paschen {\it et al.} \cite{paschen02}. 
These authors report a transition with first-order character, accompanied by hysteresis.

The atomic displacement parameters (ADP's) of Ba(2) and Ba(3) at room 
temperature were found to be large and anisotropic as previously reported 
\cite{fukuoka00,kim00}. A refinement of the low temperature patterns produced 
even larger displacements, increasing with decreasing temperature. This 
behaviour clearly indicated that the room temperature model is not adequate for 
the low temperature structure. The Ba(2) and Ba(3) sites were examined using 
difference Fourier maps. The maps present the difference between the experimental data and the model with an empty Ba site.
The results for 20K, 200K, 240K and 300K are presented in Fig.~2. 
The low temperature nuclear density distribution maps (20K) clearly show the 
tendency of Ba(2) and Ba(3) atoms to move away from the site centre. This effect 
is especially pronounced in the case of the Ba(3) site. The site separation 
increases with decreasing temperature. Inspection of the maps shows that  the 
splitting of the Ba(2) site virtually disappears above 240K but the Ba(3) site 
still shows signs of the site separation up to 300K. Multiple rattler sites were 
observed in Sr and Eu bearing type-I clathrates (X$_8$Ga$_{16}$Ge$_{30}$) but 
not in the Ba isomorph \cite{chakoumakos00,sales01,nolas00,keppens00}. This 
effect  was accommodated in the current model by splitting the Ba(2) and Ba(3) 
sites along the direction of the largest thermal displacement at all 
temperatures. The symmetry of the Ba sites was reduced as follows: Ba(2) 4a 
$\rightarrow$ 8c; Ba(3) 12b $\rightarrow$ 24e. The split Ba site occupancies 
were reduced to 0.5 to maintain the overall stoichiometry. Using this model the 
refinement produced sensible thermal displacement of Ba atoms.  At room temperature, the two models 
are almost equivalent, but, for consistency, we chose to carry out all the 
refinements using the split-site model.  Sets of refined
atomic coordinates and Debye-Waller factors at 20K and 300K are presented in Tables 1-2. An example of a refined pattern is shown in Fig.~3.  

\section{Bond geometry}
As the Ba$_6$Ge$_{25}$ crystal structure is complex, its analysis becomes easier 
by focusing on the Ge framework and on its individual building blocks (cages), 
bearing in mind that the cages share faces with each other and changes to a 
single Ge-Ge bond length affects all the cages. As already mentioned, the case 
of Ba$_6$Ge$_{25}$ is unusual as the movement of Ba atoms also affects the Ge 
framework.

\subsection{Ge framework}
All Ge-Ge bonds in the Ba$_6$Ge$_{25}$ structure are longer than in the 
elemental Ge (2.45\AA). The Ge(1)-Ge(5) (equal to 2.48\AA\ and 2.54\AA) and 
Ge(4)-Ge(4)=2.47\AA\ (the latter belongs to Ba(3) cage discussed below) 
distances are constant with temperature and are the shortest in the Ge network.  
All other Ge-Ge bonds with the exception of Ge(2)-Ge(3) and Ge(1)-Ge(6) 
monotonically increase with temperature within the range 2.535--2.615\AA\ as 
expected from thermal expansion. The Ge(2)-Ge(3) and Ge(1)-Ge(6) bond lengths 
are presented in Fig.~4 as functions of temperature. These bonds are almost 
constant below 100K and above 250K and change between 100K and 250K, with maximum 
derivative in the 200--240K temperature range, corresponding to the magnetic and 
transport anomalies \cite{paschen02}.  The transition also leads to small (less 
than one degree) changes in Ge-Ge-Ge angles associated with three-coordinated 
germaniums.

\subsection{Ba(1) cage}
The local environment of Ba(1) is presented in Fig.~5. Ba(1) is contained in a 
distorted dodecahedral cage with a 3-fold symmetry axis. Ba(1) cages share the 
pentagonal faces with other Ba(1) and Ba(3) cages and form a spiral structure in 
the crystal.
 Ba(1) together with Ge(4) and Ge(6) are located on the 3-fold axis, which is 
also the direction of their largest thermal displacement. The Ba(1)-Ba(1) 
nearest neighbour distance increases with temperature from 5.43\AA\ at 20K to 
5.46\AA\ at 300K and does not show any signs of transition. The shortest Ba(1)-
Ge(2)$\approx 3.5$\AA\ and Ba(1)-Ge(5)$\approx 3.4$\AA\ bonds are constant with 
temperature and the Ba(1)-Ge(1) shortest distance increase with temperature from 
3.42\AA\ at 20K to 3.46\AA\ at 300K. These are the shortest Ba(1)-Ge bonds in 
the Ba(1) cage. Their value suggests that Ge(1), Ge(2) and Ge(5) are at contact 
distance from Ba(1). The environment of Ba(1) seems to  be the same as in the 
type-I clathrates (the lengths are the same as calculated for 
Ba$_8$Ga$_{16}$Ge$_{30}$ cages) \cite{blake01}. The remaining Ba(1)-Ge  bonds 
change with temperature within the range 3.470--3.935\AA.

The reduction of Ge(2)-Ge(3) distance described above leads to a step like 
reduction of the Ba(1)-Ge(3) and Ba(1)-Ge(4) distances, as shown in Fig.~6. The 
increase of the Ge(1)-Ge(6) bond causes the Ge(6) atom to move away from Ba(1) 
(see Fig.~6). However, the Ba(1)-Ge(6)  increase is greater than Ba(1)-Ge(4) 
decrease so the entire Ba(1) cage expands along 3-fold axis by 0.03\AA\ over the 
20--300K temperature range. The length of the Ba(1) cage along the 3-fold axis 
is presented in Fig.~7 as a function of temperature. 

\subsection{Ba(2) cage}
The Ge(3) and Ge(6) atoms form a pseudo-cubic environment of Ba(2) presented in 
Fig.~5.
Ge(6) and Ba(2) lie on the 3-fold axis of  the unit cell and triplets of Ge(3) 
atoms on both sides of  Ba(2) form equilateral triangles rotated almost 
60$^\circ$ with respect to each other. Each Ba(2) cage is connected to six Ba(3) 
cages and together form channels in the crystal structure.  
It should be noted that only one of the split Ba sites is occupied at a time. On 
average, each of the wells is occupied with 50\% 
probability. From our data, no long-range correlation between occupied sites can 
be detected, since no superlattice Bragg reflections are observed
at low temperatures.   However, the possibility of short-range ordering, 
leading, for example, to the formation of clusters of Ba$^{2+}$ ions, could not 
be ruled out.  This aspect is currently being investigated by means of 
diffraction techniques that are sensitive to the local structure. 

The distance between the split Ba(2) sites is presented in Fig.~8 as a function 
of temperature.  The separation of Ba atoms is constant up to 100K then starts 
to decrease and rapidly falls in the 200--250K range. This sudden decrease in 
the Ba(2)-Ba(2) site distance    coincides with the kink in the lattice 
parameter (see Fig.~1). The shape of this curve closely follows the shape of the 
Ba$_6$Ge$_{25}$ resistivity curve of Paschen {\em et al.} \cite{paschen02}. At 
room temperature the split Ba(2) sites remain separated by 0.4\AA. 

The Ba(2)-Ge(3) bonds, marked with solid lines in Fig.~5, 
are constant up to 200K ($\sim$3.31\AA) and increase to 3.34\AA\ in the 200--
250K temperature range.  The almost constant value of the distance suggests a 
bonding of the three-coordinated Ge(3) with Ba(2) or a close contact of both 
atoms. These lengths are characteristic for Ba-Ge bonds observed in other Ba-Ge 
intermetallics \cite{turban73,evers80,vaughey97,kroener98,zuercher98}.  The 
Ba(2)-Ge(6) distance is too large ($>3.6$\AA) for Ge(6) to form  a bond with 
Ba(2). The length of the Ba(2) cage along the 3-fold axis is presented in Fig.~7 
as a function of temperature. Its size decreases by $\sim 0.06$\AA\ at the phase 
transition.

\subsection{Ba(3) cage}
The Ba(3) atom is contained in a heavily distorted dodecahedral cage presented 
in Fig.~9. This cage is the largest in the Ba$_6$Ge$_{25}$ structure and has 2-
fold symmetry with the rotation axis bisecting the Ge(1)-Ge(1) and Ge(4)-Ge(4) 
bonds. As in the case of Ba(2) cage, only one of the split Ba sites is occupied 
; the distance between the split sites  is shown in Fig.~8 as a function of 
temperature. The Ba(3) site separation exhibits similar behaviour to the Ba(2) 
site in the same temperature region. However the room temperature separation 
distance is equal to 0.56\AA.  Fig.~10 shows the Ba(2) site separation as a 
function of the Ba(3) site separation. The displacement amplitudes of Ba atoms 
in both cages are clearly correlated. However, the straight line fitted to the 
data does not cross the origin of the plot. Its negative offset indicates that 
Ba(3) atom can still be displaced from the cage centre while Ba(2) atom remains 
in the centre of its cage. This is consistent with the Fourier maps presented in 
Fig.~2.

The shortest Ba(3)-Ge distances are again formed by three coordinated Ge(3) and 
Ge(6) atoms. The Ba(3)-Ge(3)$\approx 3.3$\AA\  and Ba(3)-Ge(6)$\approx  
3.43$\AA\ distances are constant. The remaining Ba(3)-Ge bonds are greater than 
3.45\AA\ and vary with temperature. Once again, this suggests close contact of 
Ba and Ge atoms.

As indicated above the split site model yields acceptable temperature dependence of 
Ba thermal parameters, see Table~2. The thermal displacement of Ba(2) and Ba(3) is the largest 
among all atoms and their temperature dependence shows signs of the transition. This is most likely 
due to contraction of the Ge host  in the 200--240K temperature range, which constricts the thermal 
movement of the rattlers. Also Ge(6) exhibits a substantial but constant thermal displacement along the 3-fold axis.
This 3-coordinated Ge atom exhibits large thermal displacement because of the geometry of the lattice. 
Ge(6) is coordinated to three Ge(1) atoms but large distance to Ba(1) and Ba(2)  
along the 3-fold axis (see discussion above) allow it to move freely.

\subsection{Concluding remarks}
The introduction of the split sites in Ba(2) and Ba(3) cages leads to a minimum 
Ba(2)-Ba(3) distance. The shortest distance between Ba(2) and Ba(3) is presented 
in Fig.~11 as a function of temperature. The Ba(2)-Ba(3) closest distance is of 
the same order as in the elemental Ba (4.35\AA). It seems that Ba atoms can 
interact with each other, which is consistent with theoretical predictions 
\cite{blake01}.
\section{Discussion and conclusions}
The results of the previous sections can be summarised as follows:  we have 
observed clear changes in both lattice and internal structural parameters, 
associated with the well-known anomalies in the transport and magnetic 
properties of Ba$_6$Ge$_{25}$.  The most remarkable structural change is the 
displacement of Ba(2) and Ba(3) away from their high-symmetry site, thereby 
forming a two-fold split site occupied in a random way with 50\% probability.  
This splitting may already be present at room temperature, but is greatly 
enhanced on cooling, with a sudden increase through the transition.  We have 
also evidenced significant changes in the Ge framework at the transition.  
Interestingly, the most significant framework distortions affect the position of 
Ge(3) and Ge(6) (through the Ge(2)-Ge(3) and Ge(1)-Ge(6) bond lengths and 
associated bond angles).  Ge(3) and Ge(6) are both 3-coordinated and both form 
close-contact distances with Ba(2) and Ba(3). 
Based on this scenario, we will attempt to 
relate the observed structural changes to the known anomalies in the transport 
and magnetic properties; in particular, we will focus on three main questions: 
1) What is the driving force for the Ba-site splitting on cooling? 2) What is 
the likely effect of this distortion on the electronic structure, and is this 
sufficient to explain the observed anomalies? 3) Is there any need to go beyond 
the one-electron approximation?  Question number 3 is particularly relevant in 
the light of the spinless bipolaron mechanism proposed by Paschen {\it et al.} 
\cite{paschen02} to explain the drop in magnetic susceptibility, and of the 
observation of superconductivity in this system below 1 K \cite{grosche01}.
\subsection{Mechanism of the structural transition}
The simplest mechanism for explaining the observed behaviour of the Ba(2) and 
Ba(3) sites is that of temperature-induced de-trapping from a symmetric double-
well potential; this is a purely 'geometrical' effect, which would take place 
within a rigid framework and, in its simplest form, does not depend on the 
conduction electrons.  In this case, the close-contact interaction between Ba 
and the 3-fold-coordinated Ge atomes (Ge(3) and Ge(6) would provide both the 
attractive and the repulsive components of the potential.  A quartic potential 
bounded on both sides by infinite walls provides the simplest implementation of 
this model (Fig. 12), which can be solved numerically in both classic and 
quantum cases.  Qualitatively, the physics of this model is easy to understand 
and is in agreement with the observations:  at low temperatures, Ba(2) and Ba(3) 
are statistically trapped in one of the two wells of a symmetric double-well 
potential.  On warming, the atoms explore the available levels within each well, 
but remain confined until their energy becomes comparable with that of the 
central maximum.  When the thermal energy becomes comparable to the barrier 
height, the atoms are 'de-trapped', and become free to jump between wells and to 
occupy the central position with finite probability.  Quantitatively, the 
quartic potential model is able to explain only about 30\% of the change in Ba-Ba 
split distances, but larger changes can be obtained by using more realistic 
potentials \cite{PGRunpublished}.  Irrespectively of the details of the 
potential, the main drawback of this model is that it only produces smooth 
crossovers through the de-trapping temperature, and is therefore unable to 
describe a first-order transition as observed by Paschen {\it et al.} 
\cite{paschen02}. However, we have clearly shown that the Ge framework distorts 
in a significant way through the transition, and it is conceivable that the Ba-
Ge interaction could modify the character of the transition.  In spite of these 
difficulties, we are persuaded that the key to understand the structural 
transition is the formation of symmetric double-well potential at the Ba(2) and 
Ba(3) sites. The role of frustration in preventing a collective structural 
distortion also deserves to be investigated.

\subsection{Consequence of the structural transition on the electronic 
structure}
At low temperature the Ba(2) and Ba(3) ions are displaced; however, no
superlattice peaks are evident therefore the displaced sites must be
occupied in a random, or short-range ordered, fashion.
We would like to examine the possible effect of this on
the electronic and transport properties.
We have shown that 
structural changes occur for both the Ba sites and the Ge framework, strongly 
suggesting that the disorder will result in increased scattering for bands with 
both Ba and Ge predominant character.  This observation, by itself, is 
sufficient to explain the drop of carrier mobility on cooling, which would be a 
consequence of the reduced mean free path for the conduction electrons.  
However, the observation of anomalies in both the Seebeck coefficient and the 
magnetic susceptibility clearly indicate that changes in the electronic 
structure at the Fermi surface are taking place through the transition.  If we 
abandon the naive free-electron model and write $S$ and $\chi_e$ in the 
semiclassical one-electron model \cite{ashcroft} we obtain: \\
$S=-\frac{\pi^2}{3}\left(\frac{k_B}{e}\right)k_BT  \frac{\partial \log \sigma(E)}{\partial E} |_{E_F}$, \\
$\chi_{Pauli}=\mu_B^2 g(E_F)$\\  
where $\sigma(E)=e^2\tau(E)g(E)\frac{1}{3}v(E)^2$ is the generalised conductivity, $\tau(E)$ is the 
relaxation time, $g(E)$ is the density of states (DOS) and $v(E)$ is the 
electron velocity (which we considered isotropic for simplicity;  all terms are 
formally energy-dependent).  The observations are consistent with an overall 
{\it reduction} of the DOS and an increased {\it asymmetry} of the generalised 
conductivity through the Fermi surface.  Clearly, we would not expect the broad 
and relatively featureless bands originating from the Ge framework to display 
such behaviour.  However, recent band structure calculations by Zerec  {\em et 
al.} \cite{zerec02} have identified the presence of narrow bands with 
predominant Ba character, which cross the Fermi surface for the undistorted 
model and are significantly affected by off-site Ba displacement.  These band 
need not contribute greatly to either the overall conductivity or the carrier 
density, as speculated by Zerec  {\em et al.} \cite{zerec02}, as long as one 
still assumes that enhanced disorder is the main driving force for the resistive 
anomaly.  In summary, the observed changes in transport and magnetic properties 
can be explained by a reduced relaxation time at low temperatures and a change 
in the Ba-related bands at the Fermi surface, both effects being consistent with 
the observed structural behaviour.
\subsection{The spinless bipolaron scenario}
Based on the previous considerations, there seems to be no need for additional 
mechanisms involving strong electron-electron correlation, as proposed by 
Paschen {\it et al.} \cite{paschen02}. In particular, in the temperature-induced 
de-trapping scenario, Ba displacement would occur spontaneously, without the 
need to be associated with a single or a pair of localised electrons.  This is 
not to say that polaron physics is not relevant for this material, since there 
is clearly strong coupling between the lattice and the conduction electrons.  
However, low-temperature electronic transport associated with 'jumps' of Ba 
atoms between different wells, as proposed by Paschen {\it et al.} 
\cite{paschen02}, is highly unlikely.  This can be shown in a simple manner by 
estimating the Ba tunneling rate, $\Gamma=\Gamma_0 e^{-2\lambda}$ , where $e^{-\lambda}$ is the overlap between 
wavefunctions in the double well, $\Gamma_0$ is a typical phonon frequency and 
$\lambda \approx \frac{1}{2}\sqrt{\frac{2mV_0}{\hbar^2}}$
  \ \cite{anderson72}.  Here, $V_0$ is the barrier height and m is 
the mass of the tunnelling atom.  By setting $V_0=20$ meV (i.e., of the order of 
the transition temperature), $\Gamma_0 = 5\times10^{12}$ sec$^{-1}$ and $m= 137 m_p$ we get $\lambda=14.5$, $\Gamma=0.2$ 
sec$^{-1}$.  In other words, the tunnelling rate is macroscopically slow, as expected 
for heavy atoms such as barium.  Therefore, we believe that the low-temperature 
disorder is essentially {\it static}, and cannot be associated with electron 
hopping.
On this point, we make a final consideration: multi-well sites in type-I 
clathrates were previously associated with glass-like thermal conductivity 
\cite{chakoumakos00,sales01,nolas00,keppens00}, through a tunnel-like mechanism.  
The observation of a crystal-like thermal conductivity in Ba$_6$Ge$_{25}$, which 
displays multi-well Ba sites, calls for this proposal to be re-examined.  
Moreover, one should seriously question whether the tunnelling rates for these 
heavy-atom systems are sufficiently fast to affect low-temperature heat and 
electronic transport.
\subsection{Conclusions}
We have determined the crystal structure of the Ba$_6$Ge$_{25}$ clathrate as a 
function of temperature between 20 and 300K by neutron powder diffraction data.  
Although the structure was found to be cubic (space group P $4_1$ 3 2)at all 
temperatures, the fully-ordered structural model 
\cite{cabrera00,fukuoka00,kim00} fails to describe the data, particularly at low 
temperatures.  Much better fits are obtained by modelling the nuclear density 
for two out of the three barium positions (Ba(2) and Ba(3)) with statistically 
occupied split sites.  Ba$_6$Ge$_{25}$ was found to undergo a significant 
structural rearrangement around 240K.  Although the main effect at the 
transition is an increase of the distance between split Ba sites on cooling, we 
have observed significant changes in some of the Ge-Ge bond lengths and angles, 
indicating strong interactions between the Ba atoms and the Ge framework. The 
observed structural behaviour can be explained in a simple way by assuming that 
Ba(2) and Ba(3) are trapped, at low temperatures, within a deep double-well 
potential, from which they escape when their thermal energy becomes comparable 
to the barrier height.  From our data, the transition appears to be continuous, 
but the character of the transition is likely to be sample-dependent; other data 
\cite{paschen02} suggest a first-order character, which would require a 
significant involvement of the Ge framework in the de-trapping process.  As far 
as transport and magnetic properties are concerned, the main consequence of the 
transition is the enhanced structural disorder on cooling, which is likely to 
affect significantly all parts of the Fermi surface.  This mechanism provides a 
natural explanation for the drop in electrical conductivity through the 
transition.  The other transport and magnetic anomalies (increase in the 
absolute value of the negative Seebeck coefficient, decrease in the magnetic 
susceptibility) can be explained within the framework of the one-electron semi-
classical model, by assuming that Ba contributes with narrow bands at the Fermi 
surface, as confirmed by recent band structure calculations \cite{zerec02}.  
Electron-phonon interaction is likely to be strong in Ba$_6$Ge$_{25}$, which may 
explain the observation of superconductivity below 1 K \cite{grosche01}.  
However, we argue that Ba-site disorder is probably static at low temperatures, 
suggesting that polaron hopping involving Ba inter-well jumps is unlikely to 
contribute to the electrical or thermal conductivity. 

\section{Acknowledgements}
This work was supported in part by US DOE Office of Science grant DE-FG02-
97ER45651. Reproduction of this article, with the customary credit to the 
source, is permitted.  We are grateful to Daniel Khomskii, Alex Shluger and Laurent Chapon for 
discussing with the authors the results of this work.

\section{Figure captions}
Fig. 1 The cubic lattice parameter of Ba$_6$Ge$_{25}$ as a function of 
temperature. The error bars mark the standard deviation obtained from 
refinement. The inset shows the specific resistivity of Ba$_6$Ge$_{25}$ as a function of 
temperature.

Fig. 2 Nuclear density at the Ba(2) and Ba(3) sites for Ba$_6$Ge$_{25}$ at 20K, 
200K, 240K and 300K in difference Fourier maps determined from powder 
diffraction. The maps present the difference between the experimental data and the model with an empty Ba site.
 The size of all  maps is $5\times5$\AA$^2$.

Fig. 3 An example of a refined Ba$_6$Ge$_{25}$ diffraction pattern from the 
63$^\circ$ bank. The figure shows the observed intensities (circles), calculated 
pattern and the difference curve (solid lines). The top and bottom rows of 
tick marks indicate the positions of graphite and clathrate Bragg reflections 
respectively. The strongest line of the unidentified impurity is visible at 
3.25\AA\ of d-spacing.

Fig. 4 The Ge(1)-Ge(6) and Ge(2)-Ge(3) bond lengths in Ba$_6$Ge$_{25}$ as a 
function of temperature. The lines are guide for eye.

Fig. 5  Local environments of Ba(1) and Ba(2) sites in Ba$_6$Ge$_{25}$ at 20K 
derived from Rietveld refinement. The Ba(1) resides inside of the distorted 
dodecahedral cage (Ba atom not shown). The cage has got 3-fold symmetry with the 
Ge(4), Ge(6) and Ba(1) lying on the rotation axis. The environment of Ba(2) 
shows both split Ba(2) sites, however only one is occupied at a time. Ba(2) and  
Ge(6) are located on the 3-fold axis of the unit cell. In the case of Ba(2) site 
the lines mark the shortest Ba(2)-Ge(3) bonds.

Fig. 6 The Ba(1)-Ge(3), Ba(1)-Ge(4) and Ba(1)-Ge(6) distances in Ba$_6$Ge$_{25}$ 
as functions of temperature. The lines are guide for eye.

Fig. 7  The length of the Ba(1) and Ba(2) cages along the 3-fold axis in 
Ba$_6$Ge$_{25}$ as a function of temperature. The lines are guide for eye.

Fig. 8 The Ba(2)-Ba(2) and Ba(3)-Ba(3) split site distance in Ba$_6$Ge$_{25}$ as 
a function of temperature. The error bars represent the standard deviation 
obtained from refinement.  The lines are guide for eye.

Fig. 9   Local environment of Ba(3) site in Ba$_6$Ge$_{25}$ at 20K derived from 
Rietveld refinement. Ba(2) atoms from neighbouring cages are also shown. This 
cage has 2-fold symmetry, the rotation axis intercepts the Ge(1)-Ge(1) and 
Ge(4)-Ge(4) bonds in the middle.  Only one Ba(3) site is occupied at a time.

Fig. 10 The Ba(2)-Ba(2) split site distance as a function of Ba(3)-Ba(3) split 
site distance in Ba$_6$Ge$_{25}$. The solid line is a linear fit as discussed in 
the text.

Fig. 11 The Ba(2)-Ba(3) shortest distance in Ba$_6$Ge$_{25}$ as a function of 
temperature. The line is a guide for eye.

Fig. 12 The quatric double well potential used to demonstrate thermal de-trapping of barium atoms in double-site cages. 
The potential is bound on both sides by infinite walls, $V_0$ denotes the height of potential barrier.
\begin{table}[H]
\caption{Refined lattice constant and fractional coordinates of atoms in Ba$_6$Ge$_{25}$ at 20K and 300K.
S.G. {\em P4$_1$32}:  Ba(1) 8c (x,x,x); Ba(2) 8c (x,x,x); Ba(3) 24e (x,y,z); Ge(1) 24e (x,y,z); Ge(2) 12d (y,y+1/4,1/8); Ge(3) 24e (x,y,z); Ge(4) 8c (x,x,x); 
Ge(5) 24e (x,y,z); Ge(6) 8c (x,x,x).}
\begin{tabular}{l c c c c c c }\hline
  &\multicolumn{3}{c}{20\,K} & \multicolumn{3}{c}{300\,K}\\\cline{2-4}\cline{5-7}
Atom  & x& y& z& x & y& z \\\hline
Ba(1) & 0.0642(4) & 0.0642(4) & 0.0642(4) & 0.0618(3) & 0.0618(3) & 0.0618(3) \\
Ba(2) & 0.3606(6) & 0.3606(6) & 0.3606(6) & 0.3670(8) & 0.3670(8) & 0.3670(8) \\
Ba(3) & 0.1867(6) & 0.4418(7) & 0.1539(5) & 0.1893(6) & 0.4404(7) & 0.1443(5) \\
Ge(1) & 1.0005(2) & 0.2974(1) & 0.0428(2) & 0.9988(2) & 0.2970(1) & 0.0419(1) \\
Ge(2) & 0.8320(2) & 0.0820(2) & 1/8               & 0.8307(1) & 0.0807(1) & 1/8 \\
Ge(3) & 0.8534(2) & 0.9142(2) & 0.0836(2) & 0.8520(1) & 0.9153(2) & 0.0834(2) \\
Ge(4) & 0.9240(2) & 0.9240(2) & 0.9240(2) & 0.9240(2) & 0.9240(2) & 0.9240(2) \\
Ge(5) & 0.1270(2) & 0.2588(2) & 0.9350(2) & 0.1264(2) & 0.2597(2) & 0.9345(2) \\
Ge(6) & 0.2169(2) & 0.2169(2) & 0.2169(2) & 0.2181(2) & 0.2181(2) & 0.2181(2) \\\hline
 & a [\AA] & R$_{wp}$ & R$_{p}$ & a [\AA] & R$_{wp}$ & R$_{p}$ \\
 & 14.52828(8) & 0.0651& 0.0641 & 14.54536(7) & 0.0524 & 0.0508\\
\hline
\end{tabular}
\end{table}
\begin{table}[H]
\caption{Refined thermal parameters of atoms in Ba$_6$Ge$_{25}$ at 20K and 300K in [\AA$^2$]. 
Ba(2) and Ba(3) thermal factors were refined using an isotropic model (U$_{iso}$=U$_{11}$).}
\begin{tabular}{r c r r r r r r}
 \hline
T [K] & Atom & 100U$_{11}$ & 100U$_{22}$ & 100U$_{33}$ & 100U$_{12}$ & 100U$_{13}$ &  100U$_{23}$ \\ \hline
         20 & Ba(1) &      1.0(1)&      1.0(1)&      1.0(1)&      1.3(2)&      1.3(2)&      1.3(2) \\
         20 & Ba(2) &      2.5(4)&          2.5&          2.5&          0.0&          0.0&          0.0 \\
         20 & Ba(3) &      1.8(2)&          1.8&          1.8&          0.0&          0.0&          0.0 \\
         20 & Ge(1) &     -0.2(1)&      0.2(2)&      1.7(2)&     -0.4(1)&      0.5(1)&     -0.5(1) \\
         20 & Ge(2) &      2.1(2)&      2.1(2)&      3.3(3)&      1.0(3)&      1.0(2)&     -1.0(2) \\
         20 & Ge(3) &     -1.4(1)&      1.9(2)&      1.0(1)&     -0.4(1)&     -0.1(1)&      0.7(1) \\
         20 & Ge(4) &      0.0(1)&      0.0(1)&      0.0(1)&      0.3(1)&      0.3(1)&      0.3(1) \\
         20 & Ge(5) &     -0.2(1)&     -0.2(2)&      1.1(2)&     -0.2(1)&      0.1(1)&      0.6(1) \\
         20 & Ge(6) &      2.5(2)&      2.5(2)&      2.5(2)&      0.5(2)&     -0.5(2)&     -0.5(2) \\
       300 & Ba(1) &      1.3(1)&      1.3(1)&      1.3(1)&      0.9(2)&      0.9(2)&      0.9(2) \\
       300 & Ba(2) &      3.3(4)&          3.3&          3.3&          0.0&          0.0&          0.0 \\
       300 & Ba(3) &      2.2(2)&          2.2&          2.2&          0.0&          0.0&          0.0 \\
       300 & Ge(1) &     -0.1(1)&      1.3(2)&      1.3(1)&      0.4(1)&      0.3(1)&     -0.1(1) \\
       300 & Ge(2) &      1.0(1)&      1.0(1)&      1.0(2)&      0.7(2)&      0.1(1)&     -0.1(1) \\
       300 & Ge(3) &      0.3(1)&      0.7(1)&      1.7(1)&     -0.5(1)&       0.31(9)&      0.1(1) \\
       300 & Ge(4) &       0.73(9)&       0.73(9)&       0.73(9)&      0.0(1)&      0.0(1)&      0.0(1) \\
       300 & Ge(5) &      0.7(1)&      0.5(1)&      1.2(1)&      -0.10(9)&     -0.1(1)&      0.4(1) \\
       300 & Ge(6) &      2.6(1)&      2.6(1)&      2.6(1)&      0.4(1)&     -0.4(1)&     -0.4(1) \\\hline
\end{tabular}
\end{table}
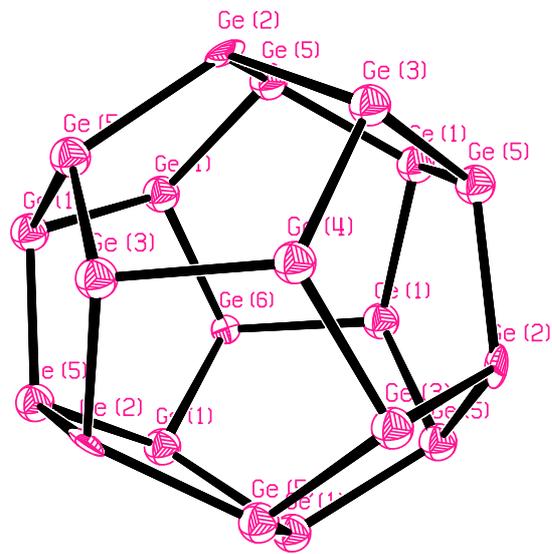
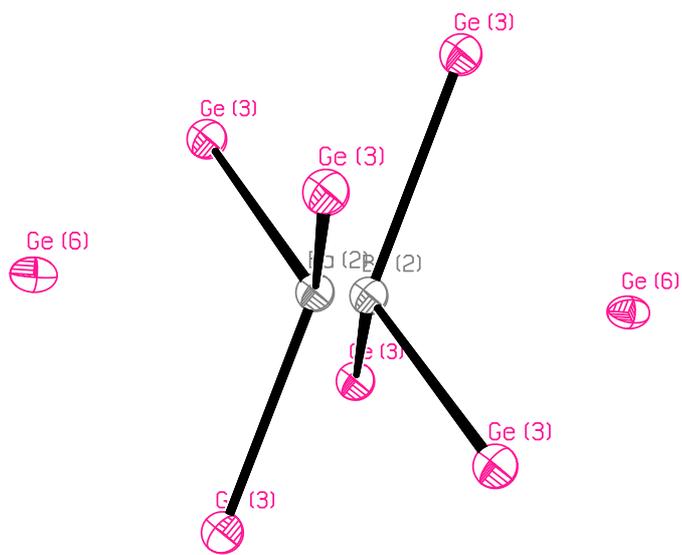
\end{document}